\documentclass[review]{elsarticle}
\usepackage[margin=1in]{geometry}
\usepackage{float}
\usepackage[utf8]{inputenc}
\usepackage[T1]{fontenc}
\usepackage{csquotes}
\usepackage[english]{babel}
\usepackage{textcomp}

\usepackage{lineno}
\usepackage{import}
\modulolinenumbers[1]

\journal{Fuel}

\usepackage{color,soul}

\usepackage{todonotes}
\usepackage{gensymb}
\usepackage{caption}
\usepackage{csvsimple}
\usepackage{booktabs, tabularx, wrapfig}
\usepackage{array, ltablex, multirow}
\usepackage{placeins}
\usepackage{graphicx}
\usepackage{subcaption}

\usepackage{siunitx}
\usepackage[version=4]{mhchem}

\usepackage{threeparttable}
\usepackage{adjustbox}
\usepackage{diagbox}
\usepackage{rotating}
\usepackage{siunitx}
\usepackage{graphicx}
\usepackage{url}
\usepackage[version=4]{mhchem} 
\usepackage[doublespacing]{setspace}
\usepackage{csvsimple}
\usepackage{makecell}
\usepackage[acronym, style = super, nonumberlist, nogroupskip]{glossaries}
\makeglossaries

\newacronym{ML}{ML}{Machine Learning}
\newacronym{TPOT}{TPOT}{Tree-based Pipeline Optimization Tool}
\newacronym{MAPE}{MAPE}{Mean Absolute Percentage Error}
\newacronym{MAE}{MAE}{Mean Absolute Error}
\newacronym{RMSE}{RMSE}{Root Mean Squared Error}
\newacronym{LC/1H NMR}{LC/1H NMR}{Liquid Chromatography/1H Nuclear Magnetic Resonance}
\newacronym{FTIR}{FTIR}{Fourier Transform Infrared}
\newacronym{IR}{IR}{Infrared}
\newacronym{NIR}{NIR}{Near-Infrared}
\newacronym{MIR}{MIR}{Mid-Infrared}
\newacronym{GC-MS}{GC-MS}{Gas Chromatography with Mass Selective Detection}
\newacronym{PLS}{PLS}{Partial Least Squares}
\newacronym{RM}{RM}{Raman Spectroscopy}
\newacronym{GLM}{GLM}{Generalized Linear Model}
\newacronym{NMF}{NMF}{Non-negative Matrix Factorization}
\newacronym{PCA}{PCA}{Principal Component Analysis}
\newacronym{NMR}{NMR}{Nuclear Magnetic Resonance}
\newacronym{GC}{GC}{Gas Chromatography}
\newacronym{MS}{MS}{Mass Spectrometry}
\newacronym{ASTM}{ASTM}{American Society for Testing and Materials}
\newacronym{DMCO}{DMCO}{1,4-dimethylcyclooctane}
\newacronym{ATR}{ATR}{Attenuated Total Reflectance}

\begin{document}

    
    \begin{frontmatter}
    
    \title{A Structured Framework for Predicting Sustainable Aviation Fuel Properties using Liquid-Phase FTIR and Machine Learning}
    
    \author[1]{Ana E. Comesana}
    \author[1]{Sharon Chen}
    \author[2]{Kyle E. Niemeyer}
    \author[1]{Vi H. Rapp\corref{cor1}}
    
    \address[1]{Energy Technologies Area, Lawrence Berkeley National Laboratory, Berkeley CA, United States}
    \address[2]{School of Mechanical, Industrial and Manufacturing Engineering, Oregon State University, Corvallis OR, United States}
    \cortext[cor1]{Corresponding author: vhrapp@lbl.gov}
    
    \begin{abstract}
    Sustainable aviation fuels have the potential for reducing emissions and environmental impact.
    To help identify viable sustainable aviation fuels and accelerate research, several machine learning models have been developed to predict relevant physiochemical properties.
    However, many of the models have limited applicability, leverage data from complex analytical techniques with confined spectral ranges, or use feature decomposition methods that have limited interpretability.
    Using liquid-phase Fourier Transform Infrared (FTIR) spectra, this study presents a structured method for creating accurate and interpretable property prediction models for neat molecules, aviation fuels, and blends. 
    Liquid-phase FTIR spectra measurements can be collected quickly and consistently, offering high reliability, sensitivity, and component specificity using less than 2 mL of sample.
    The method first decomposes FTIR spectra into fundamental building blocks using  Non-negative Matrix Factorization (NMF) to enable scientific analysis of FTIR spectra attributes and fuel properties.
    The NMF features are then used to create five ensemble models for predicting final boiling point, flash point, freezing point, density at 15 \degree C, and kinematic viscosity at -20 \degree C. 
    All models were trained using experimental property data from neat molecules, aviation fuels, and blends.
    The models accurately predict properties while enabling interpretation of relationships between compositional elements of a fuel, such as functional groups or chemical classes, and its properties.
    To support sustainable aviation fuel research and development, the models and data are available on an interactive web tool.
    
    \end{abstract}
    
    
    \end{frontmatter}


    \printglossary[type=\acronymtype,title={Nomenclature}]
    
    
\section{Introduction}

Bio-derived jet fuel research has gained traction as a way to mitigate climate change and reduce emissions.
However, high cost and high-volume requirements often delay experimental property testing of novel blends for years after initial bench-scale demonstrations.
This increases the risk of scaling-up bio-jet fuels that do not perform as well as expected~\cite{balakrishnan2015}.
Simple models to calculate the properties of bio-jet fuels and blends are used to reduce risks, but these models often use linear-by-volume or linear-by-molar mass calculations that limit their applicability~\cite{boppadati_2024,wang_2019,boppadati_2023}.

In the last decade, machine learning has been used to develop property prediction models that are more accurate and can help accelerate bio-jet development~\cite{comesana2022,wang_2019,boppadati_2024,boppadati_2023,wang_2021,yang_2021,johnson_2006,mdli_2021,mdsaldana_2011,mdsaldana_2013,st_john_2017}.  
Several machine learning models predict chemical and physical properties of molecules using chemical descriptors or functional group counts~\cite{mdespinosa_2001,mddai_2013,mdgakh_1994,mdcherqaoui_1994,mdkarthikeyan_2005,mdgharagheizi_2008_fp,mdmodarresi_2006,mdbergstrom_2003,mdli_2021,mdsaldana_2013,mdsola_2008,mdroubehie_fissa_2019,mdkatritzky_1996,mdzhokhova_2003}.
Because chemical descriptors and functional group counts require knowledge of the exact chemical composition of the blend, these models are unable to predict properties of fossil and bio-derived fuels and fuel blends. 
As a result, researchers developed machine learning models that use different analytical techniques to characterize fuels, such as \gls{NMR}, \gls{GC}, and \gls{MS}~\cite{aslani2022,caswell_1989,johnson_2006,yang_2021}.
These analytical techniques can characterize neat molecules and blends by providing insights into the functional groups, bonds, and component classes present. 
Many techniques are also able to differentiate between isomers and provide fundamental chemical information about a fuel's components that can be used as inputs for machine learning models~\cite{aslani2022}. 
For example, one study used \gls{LC/1H NMR} with multiple regression analysis to develop a group property approach, similar to functional group theory, for predicting properties of distillate fuels, including flash point, density, and final boiling point~\cite{caswell_1989}. 
Another study predicted similar properties using partial least squares with \gls{NIR} spectroscopy, \gls{RM}, and \gls{GC-MS}~\cite{johnson_2006}.
Two-dimensional \gls{GC} and Monte Carlo sampling have also been used to predict properties of sustainable aviation fuels~\cite{yang_2021}.

\gls{FTIR} spectroscopy is another popular technique for developing predictive models because it offers rapid data acquisition, component specificity, and sensitivity~\cite{Bradley_2021,wang_2019,wang_2021,boppadati_2024,boppadati_2023,fodor_1999,Daly2016,Daly2019,yang_2022,moro_2020,riverabarrera_2020}.
Many researchers have developed property prediction models using a limited range of vapor-phase \gls{FTIR} spectra.
For example, researchers have used a limited region of vapor-phase \gls{FTIR} spectra with LASSO-regularized linear models to predict physical and chemical properties including molecular weight, density, and initial boiling point of fuels~\cite{wang_2019}. 
Expanding on this work, they later added functional group theory information for a selected region of the full \gls{FTIR} spectra to their LASSO-regularized models to predict the same properties for hydrocarbon fuels~\cite{wang_2021}.
Other researchers used a limited region of vapor-phase \gls{FTIR} spectra and Elastic-net regularized linear models to predict similar properties of pure hydrocarbons~\cite{boppadati_2024}. 
In addition, liquid-phase \gls{FTIR} spectra has been used with partial least squares models to determine cetane number, net heat of combustion, viscosity, carbon-to-hydrogen ratio, and density of middle distillate fuels~\cite{fodor_1999}.

Although the literature presents well-performing models for predicting various fuel properties, the methods have several limitations~\cite{Bradley_2021,wang_2019,wang_2021,boppadati_2024,boppadati_2023,fodor_1999,Daly2016,Daly2019,yang_2022,moro_2020,riverabarrera_2020}.
For example, some studies developed predictive models using narrow \gls{FTIR} spectra ranges, which may omit important information necessary for accurately predicting a variety of fuel properties~\cite{boppadati_2023,boppadati_2024,wang_2019, wang_2021}.
Additionally, measuring vapor-phase \gls{FTIR} spectra is time consuming and complicated since any variations in temperature or pressure in the sample lines or gas cell will result in significant error ~\cite{Stec2011}.
Further, many of the models do not provide an easily interpretable analysis of the spectral features (e.g., how much a feature contributes to the predictions) and use non-standard error metrics, making it difficult to assess the model's performance on unseen data~\cite{wang_2019,boppadati_2024,boppadati_2023} 

The purpose of this study is to develop a structured approach for creating accurate and interpretable property prediction models using liquid-phase \gls{FTIR} spectra of neat molecules, aviation fuels, and their blends.
Liquid-phase \gls{FTIR} spectra are used because they can be collected quickly and consistently, requiring only a few drops of fuel.
The structured approach uses \gls{NMF} to deconstruct \gls{FTIR} spectra into fundamental building blocks that enable scientific analysis of spectral characteristics and fuel properties.
Because of its non-negativity constraints, \gls{NMF} is well-suited for spectra data and can create meaningful, additive features that are easy to interpret. 
Using the \gls{NMF} features, we developed five ensemble models for predicting final boiling point, flash point, freezing point, density at \qty{15}{\degreeCelsius}, and kinematic viscosity at \qty{-20}{\degreeCelsius}. 
We trained the models using experimental property data from neat molecules, aviation fuels, and blends.
In the following sections, this paper outlines the methods for creating features from \gls{FTIR} spectra using \gls{NMF} and developing models with these features to predict fuel properties.
Next, model performance metrics are presented, along with an analysis and interpretation of the features used in the models.
Conclusions and future recommendations are also provided.

\section{Methods}\label{Sec:meth}

\subsection{Experimental Property Data}\label{Sec:meth-propdata}

Experimental property data for neat molecules, aviation fuels, and blends were collected from multiple published sources~\cite{pubchem,chemspider,springermaterials,saldana_data,bradley_data,james_2017,rosenkoetter,rolland_garcia_2014,yang2019_2,optima,yaw_2015,comesana2022}.
Data includes alkanes, isoalkanes, cylcoalkanes, aromatics, alcohols, conventional fossil-based jet fuels (e.g., Jet-A, F-24), \gls{ASTM} approved alternative aviation fuels (e.g., HEFA, ATJ, SPK, and farnesane), and other promising aviation molecules, including \gls{DMCO}, pinane, and limonene.
Property data for F-24 was provided by the supplier, while a local certification lab measured properties for Jet A, HEFA, and their blends.
Properties for reference fuel blends that include n-heptane and isooctane were calculated using the linear-by-volume method.
Reference fuel blends with less than 10\% ethanol were also calculated using the linear-by-volume method.
Table~\ref{Tab:tbl_num} shows the number of neat molecules, fuels and fuel blends with experimental data for each property.
Data for the 87 neat molecules, aviation fuels, and blends used in this study can be found in Tables S3 and S4 and on the Feedstock to Function Website: \url{feedstock-to-function.lbl.gov}.

\begin{table}[htb]
    \caption{Number of samples with experimental data for each of the properties. Note that density is measured at \qty{15}{\celsius} and kinematic viscosity is measured at \qty{-20}{\celsius} }
    \label{Tab:tbl_num}
    \centering
    \begin{tabular}{@{}lccccc@{}}
    \toprule
                        & Final boiling point & Flash point & Freezing point & Density & Kinematic Viscosity \\
    \midrule
    Neat molecules          & 42  & 29 & 40 & 38 & 17 \\
    Aviation Fuels                   & 4 & 4 & 4 & 5 & 4 \\
    Blends                  & 26 & 26 & 26 & 26  & 10  \\
 
    \bottomrule

\end{tabular}
\end{table}

\subsection{Liquid-phase FTIR Spectra Measurements}\label{Sec:meth-specdata}

We collected \gls{FTIR} spectra of 87 neat molecules, aviation fuels, and blends from published literature (67) and measurements (24) using the method described by Daly et al.~\cite{Daly2016,Daly2019} with minor modifications. 
We modified the method to ensure spectra collected were consistent, repeatable, and matched spectra from previous literature~\cite{Daly2016,Daly2019}.
Specifically, spectra of viable bio-jet molecules, jet fuels, and jet fuel blends were collected using a Thermo Scientific Nicolet iS50 \gls{FTIR} spectrometer located at Berkeley Lab's Molecular Foundry with a Smart iTR Attenuated Total Reflectance sampling accessory containing diamond crystal with ZnSe lens.
Viton square-profile o-rings with 316 stainless steel tags prevented the liquid fuels from spreading and evaporating during spectra measurements.
Spectra were collected between \qty{4000}{\cm^{-1}} and \qty{650}{\cm^{-1}} at a \qty{2}{\cm^{-1}} resolution.
Further details about the fuels and spectra measurements are provided in Section S-1 of the Supplemental Information (SI).

\subsection{Data Prepossessing and Dimensionality Reduction}\label{Sec:meth-specclean}

\gls{FTIR} spectra required preprocessing to reduce noise and variability between measurements before it could be used to train machine learning models~\cite{fodor_1999,johnson_2006,zhang_2020}.
We used Python libraries (Pandas and NumPy) to bin, smooth, shift, clip, and normalize the spectra~\cite{pandas-scipy-2010,numpy}.
To remove small variations in spectra wavelengths between \gls{FTIR} spectra measurements, we coalesced the data into bins.
For example, creating a bin at \qtyrange{650.4}{649.9}{\cm^{-1}} combined the published data sampled at \qty{650.162}{\cm^{-1}} and measured data sampled at \qty{650.145}{\cm^{-1}} so it could be input into the machine learning model.
We then implemented a moving average with a window width of 15 bins to smooth experimental noise in the spectra while still retaining the shape of major peaks.
To remove noise caused by mechanical vibrations or other external factors, we uniformly shifted the baselines of spectra by the mean of the values between \qty{4000}{\cm^{-1}} and \qty{3600}{\cm^{-1}}~\cite{johnson_2006,zhang_2020}.
We selected this range because it did not contain absorption bands for jet fuels or compounds of interest in this study, providing a good proxy for background noise. 
Negative values were clipped to zero and the data between \qty{2500}{\cm^{-1}} and \qty{2000}{\cm^{-1}} were removed because diamond crystals cause absorption bands in this region.
We normalized the data by their sum to ensure that peak magnitudes for all spectra were comparable and reduce instrumental error.
Equation~\eqref{eq:norm} shows the calculation used to normalize a spectra, with $X_{wn}$ representing the spectra of interest at wave number \textit{wn} and $X_i$ representing the spectra at all possible wave numbers, $i$:
\begin{equation}
\text{Normalized } X_{wn} = \frac{X_{wn}}{\sum_{i=\max}^{i=\min} X_i}
\label{eq:norm}
\end{equation}
The final shape and magnitude of the normalized spectra used for model training reflect the proportions of bonds, functional groups, and component classes~\cite{wang_2019}.
After preprocessing, scikit-learn's \gls{NMF} function is used to reduce dimensionality and extract features for the model~\cite{scikit-learn_2011}.
Specifically, \gls{NMF} decomposes the spectra data into the product of two lower-rank matrices and iteratively modifies the values within the matrices to create the best approximation of the whole dataset~\cite{paatero1994}. 
\gls{NMF} is a non-convex optimization problem with no closed-form solution~\cite{lee1999,lee2000}.
The process starts by randomly initializing two matrices, $\mathbf{A}$ and $\mathbf{B}$, that serve to factorize the spectra dataset.
$\mathbf{A}$ is the coefficient matrix, representing the original spectra samples as additive linear combinations of features in $\mathbf{B}$, which
is the basis matrix containing the non-negative features of the data.
The matrices $\mathbf{A}$ and $\mathbf{B}$ are then iteratively updated to minimize the reconstruction error between $\mathbf{A} \times \mathbf{B}$ and the original dataset.
This process decomposes the spectra data into features that preserve spectral characteristics, making them easy to interpret~\cite{lee1999}.
Further, the features can be used to reconstruct all original spectra in the database and are used for the models.

We split the data into training (68\%), validation (12\%), and testing (20\%) sets, and then fit the \gls{NMF} models using scikit-learn~\cite{scikit-learn_2011}.
We used the validation set for hyperparameter tuning and for selecting the final models.
The testing set was used to estimate the error of the whole pipeline on unseen data.
To expand the applicability domain for \gls{NMF}, we added spectra of molecules and blends relevant to alternative aviation fuel to the training data set, even though they did not have published experimental property data.
We fit \gls{NMF} models with up to 30 components for each property, and then selected the number of components that yielded the best performance on the validation set.
We also found that randomizing \gls{NMF} did not highly affect the components created or the number of components selected for the models.  

\subsection{Model Development Using Machine Learning}
\begin{figure}[htb]
    \centering
    \includegraphics[width=0.9\textwidth]{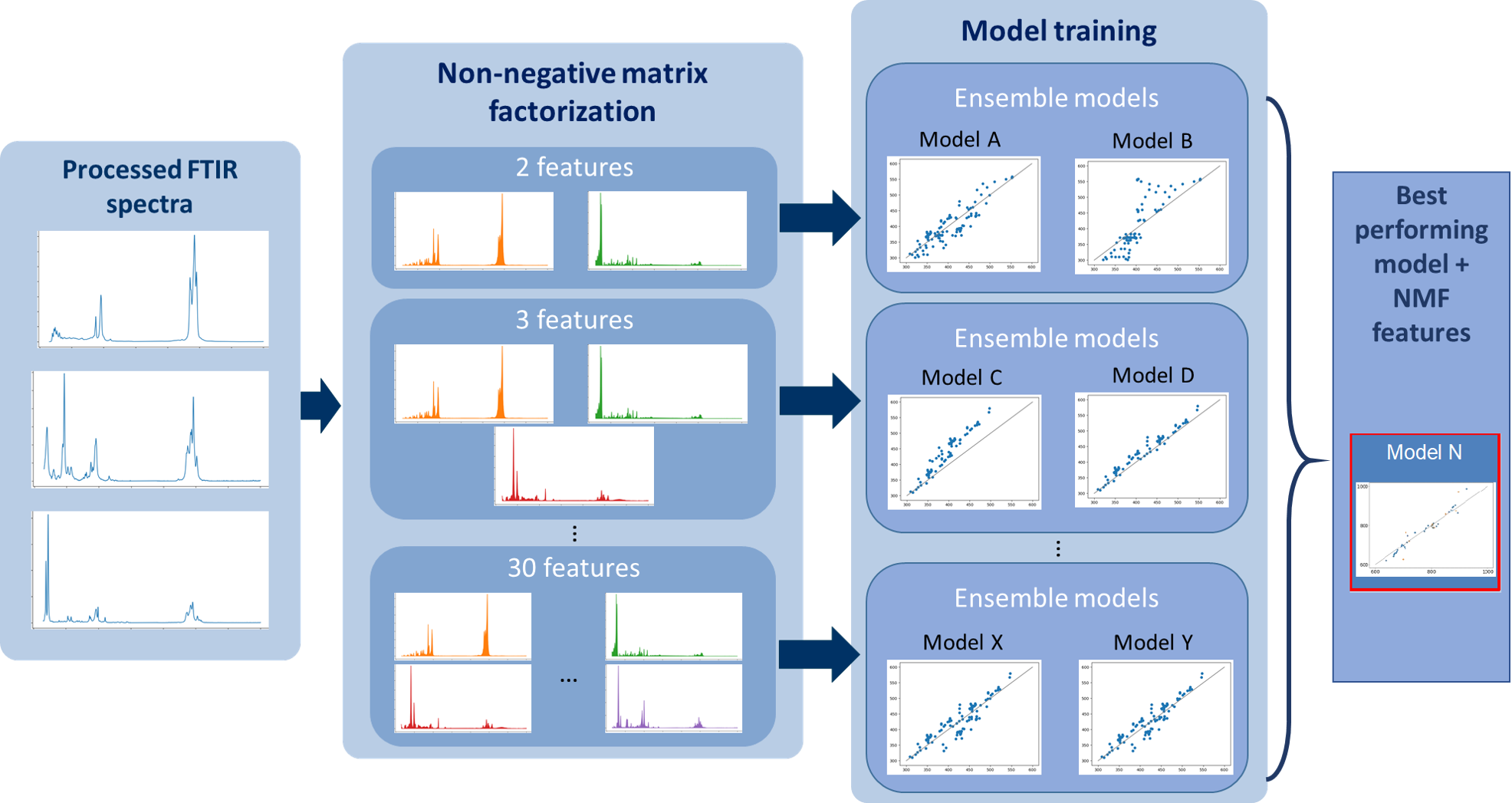}
    \caption{Schematic of the model development process}
    \label{fig:pipeline}
\end{figure}

To train machine learning models, we used \gls{NMF} to create groups of 2 to 30 features.
For kinematic viscosity, groups of up to 15 features were created to account for the smaller dataset size and avoid overfitting.
Extra trees and random forest regressors were then fit to the training set with different hyperparameters.
We selected these regressors because of their ability to provide prediction intervals with quantile regression forests or quantile extra trees, which helped us quantify prediction errors~\cite{meinshausen_2006}.
The ensemble regressors were trained using scikit-learn~\cite{scikit-learn_2011}.
The validation set was used to optimize the number of components, model architecture, and hyperparameters.
For each property, we picked the number of features and model architecture that yielded the best performance on the validation set. 
We then use the test set to estimate the model's error on unseen data.
Figure~\ref{fig:pipeline} shows a schematic of the model development process.

\label{Sec:meth-model}

\section{Results and Discussion}\label{Sec:res}
\subsection{Model Performance}
\begin{figure}[htb!]
    \begin{subfigure}{0.33\textwidth}
    \includegraphics[width=\textwidth]{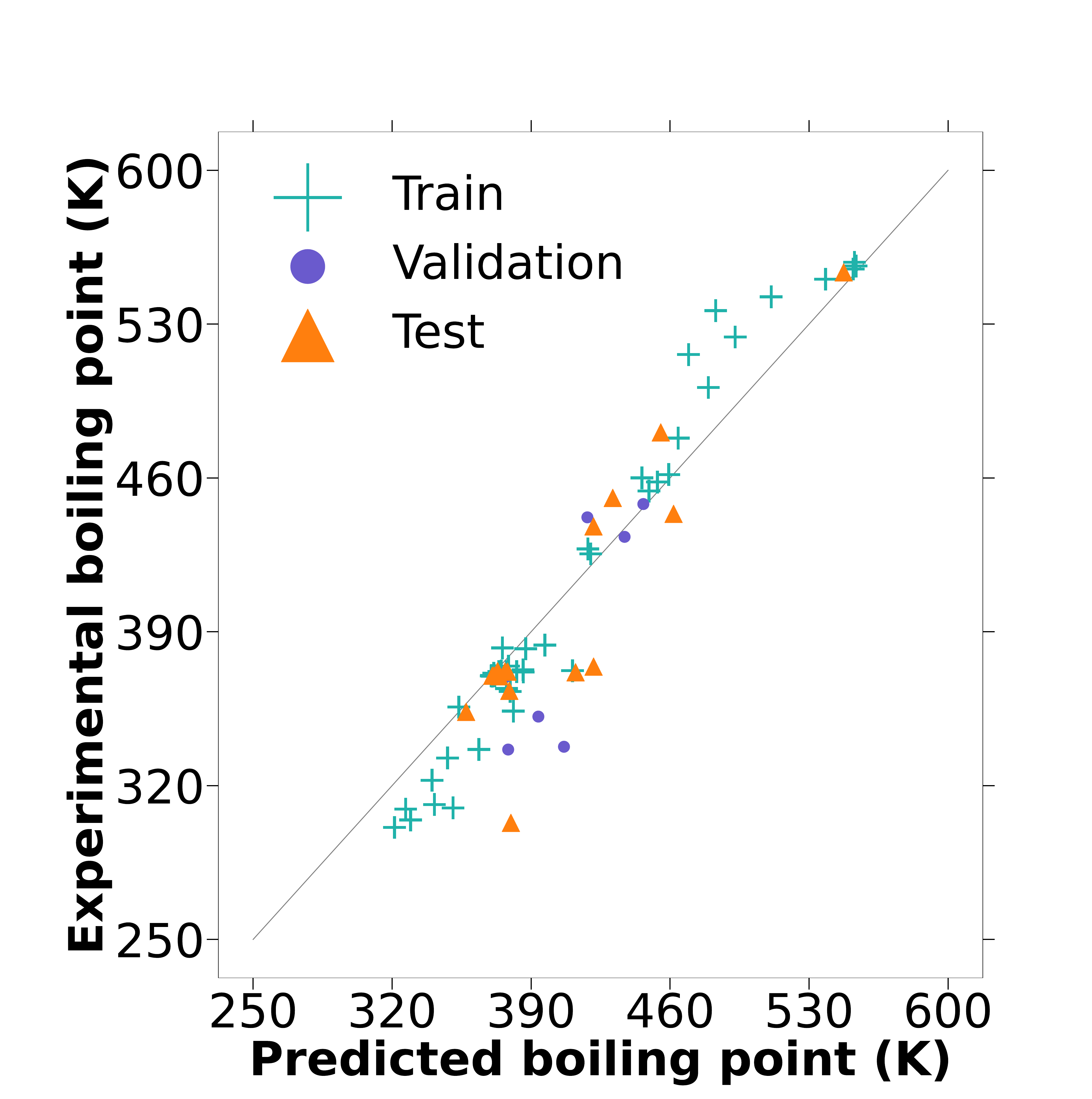}
    \caption{final boiling point}
    \label{fig:bp_par}
    \end{subfigure}
    \begin{subfigure}[b]{0.33\textwidth}
        \includegraphics[width=\textwidth]{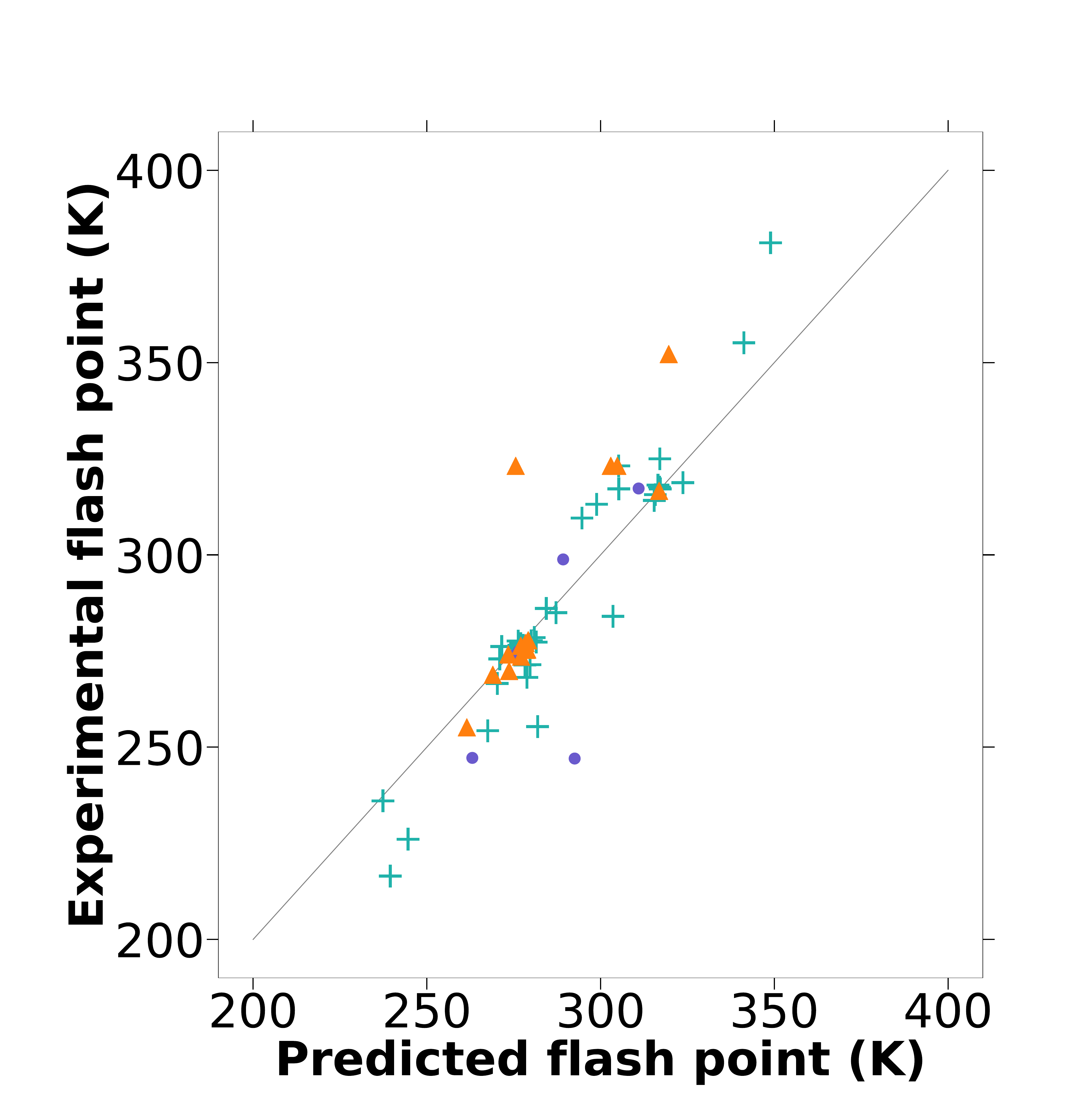}
        \caption{flash point}
        \label{fig:fp_par}
    \end{subfigure}
    \begin{subfigure}{0.33\textwidth}
    \centering
        \includegraphics[width=\textwidth]{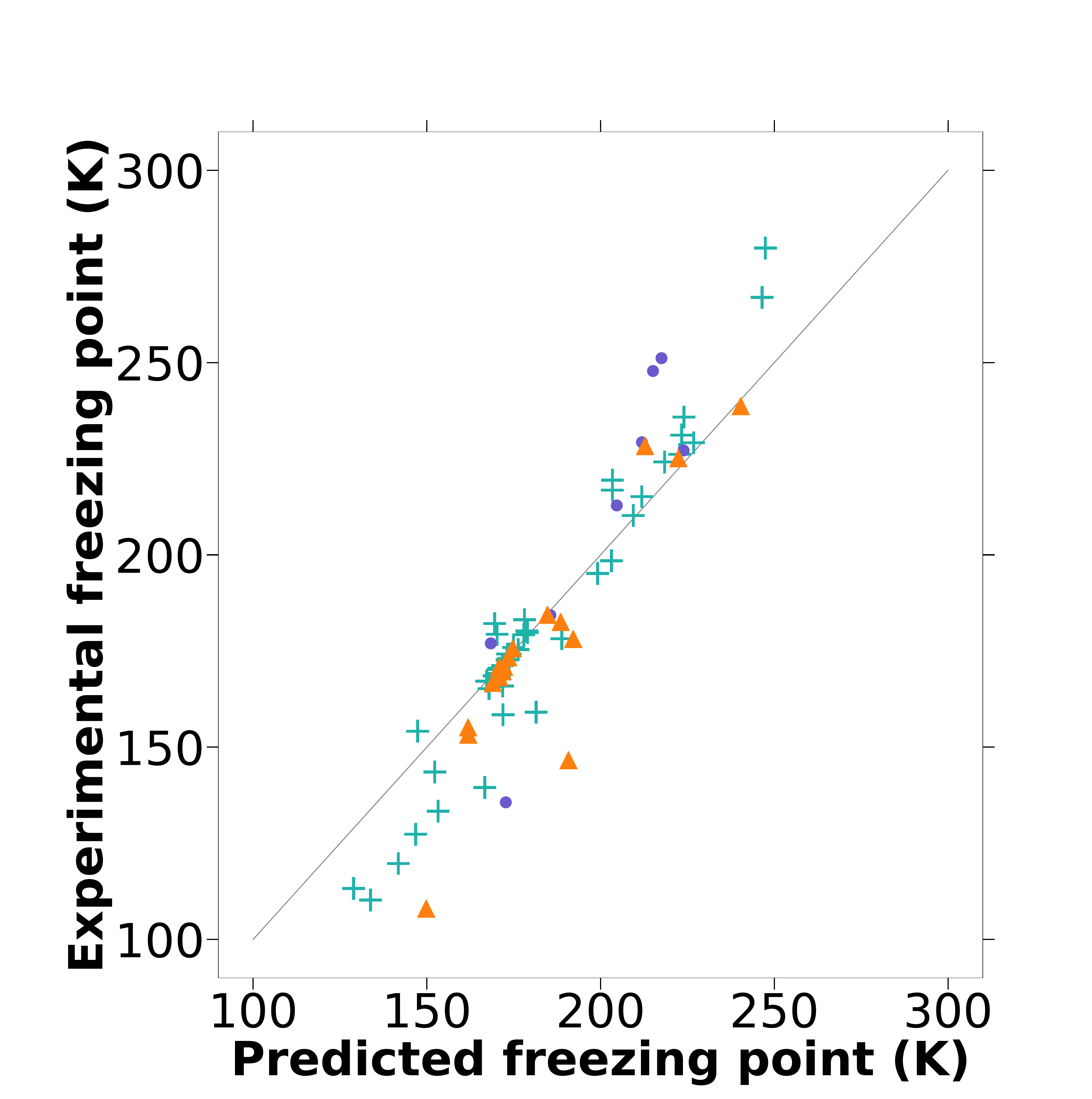}
        \caption{freezing point}
        \label{fig:mp_par}
    \end{subfigure}

    \hspace{0.1515\textwidth}
        \begin{subfigure}{0.33\textwidth}
        \centering
        \includegraphics[width=\textwidth]{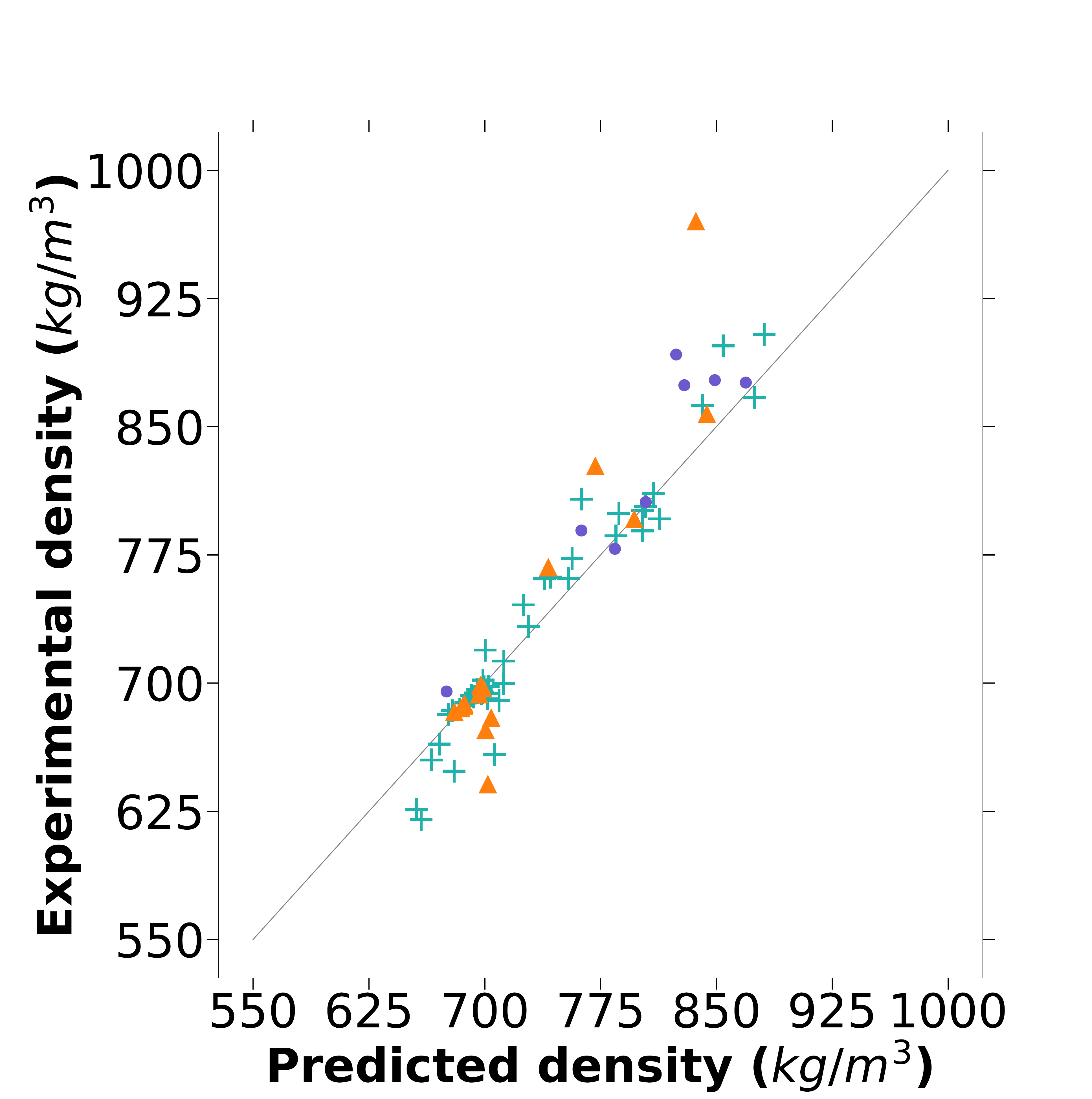}
        \caption{density at \qty{15}{\degreeCelsius}}
        \label{fig:den_par}
    \end{subfigure}
    \begin{subfigure}{0.33\textwidth}
    \centering
        \includegraphics[width=\textwidth]{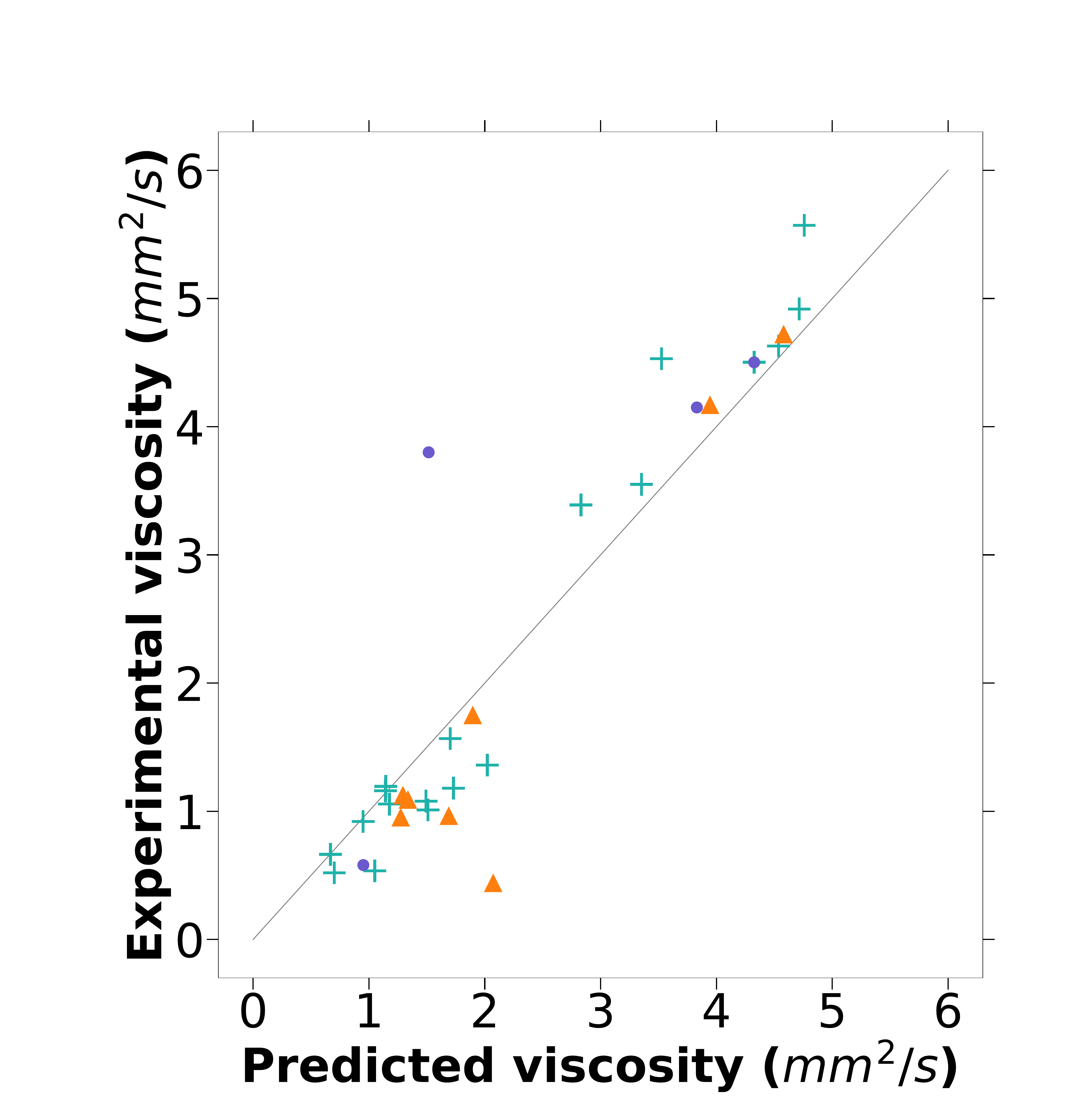}
        \caption{kinematic viscosity at \qty{-20}{\degreeCelsius}}
        \label{fig:vic_par}
    \end{subfigure}

    \caption{Parity plots showing experimental and predicted values of all models, split by training, validation, and testing sets.}
    \label{fig:all_parity}
\end{figure}

Using the method described in Section~\ref{Sec:meth-model}, we developed predictive models for final boiling point, flash point, freezing point, kinematic viscosity, and density of neat molecules, aviation fuels, and blends.
Table~\ref{Tab:model_perf} summarizes each model's characteristics and test performance, while Figure~\ref{fig:all_parity} shows parity plots for the models.
Additional performance parameters such as training errors and overall model errors are provided in Section S-2 of the SI.

All predictive models perform consistently with other spectra-based property prediction models even though the dataset used in this study was more diverse, including a larger range of properties values, molecule types, and blends~\cite{caswell_1989,johnson_2006,wang_2021,boppadati_2024,fodor_1999,yang_2022,moro_2020,riverabarrera_2020}. 
Further, the results show that the models are capable of accurately predicting properties when trained on clustered data.
For example, the test set \gls{MAE} for the viscosity model is \qty{0.45}{\mm^2\per\s} despite exhibiting two clusters between 0 and 2 and above 3.6 (Figure~\ref{fig:all_parity}e).
The viscosity model could be improved with additional experimental data, particularly between the two clusters. 
\gls{MAPE} does not seem to be an appropriate performance metric for kinematic viscosity, since more than one quarter of the viscosity data falls between 0 and 1, which inflates this metric due to division by smaller values (\gls{MAPE} = mean(|experimental $-$ predicted|/experimental)).

\begin{table}[htb]
    \caption{Characteristics and performance for the five property prediction models. Density is measured at \qty{15}{\degreeCelsius} and viscosity is measured at \qty{-20}{\degreeCelsius}.}
    \label{Tab:model_perf}
    \centering
    \begin{threeparttable}
    \begin{tabular}{@{}lccccc@{}}
        \toprule
                            & Final boiling point & Flash point & Freezing point & Density & Kinematic Viscosity \\
        & (\unit{\kelvin}) & (\unit{\kelvin}) & (\unit{\kelvin}) & (\unit{\kg\per\m^3}) & (\unit{\mm^2\per\s}) \\
        \midrule
        Test \gls{MAE}                 & \num{17.17} & \num{9.6} & \num{8.9} &  \num{22.2} &\num{0.45} \\
        Test RMSE**                & \num{26.39} & \num{16.7}  & \num{15.9} &  \num{41} & \num{0.66} \\
        Test \gls{MAPE}                & 4.6\%   & 3\%                  & 6.1\%                 & 2.7\% & 66.8$^*$\% \\
        Number of features  & 16      & 8                     & 15                   & 9 & 10 \\
        Number of compounds & 72    & 59                   & 70                  & 69 & 31 \\ 
        Data range & \numrange{300}{559} & \numrange{216}{382} & \numrange{108}{278} & \numrange{620}{971} & \numrange{0.4}{5.6} \\
        
        \bottomrule
    \end{tabular}
    \begin{tablenotes}\footnotesize
    \item[*] Most viscosity data lies between 0 and 2.
    \item[**] Root Mean Squared Error.
    \end{tablenotes}
    \end{threeparttable}
\end{table}

\subsection{Model Features}
\begin{figure}[btp]
    \begin{subfigure}{0.33\textwidth}
    \centering
    \includegraphics[width=\textwidth]{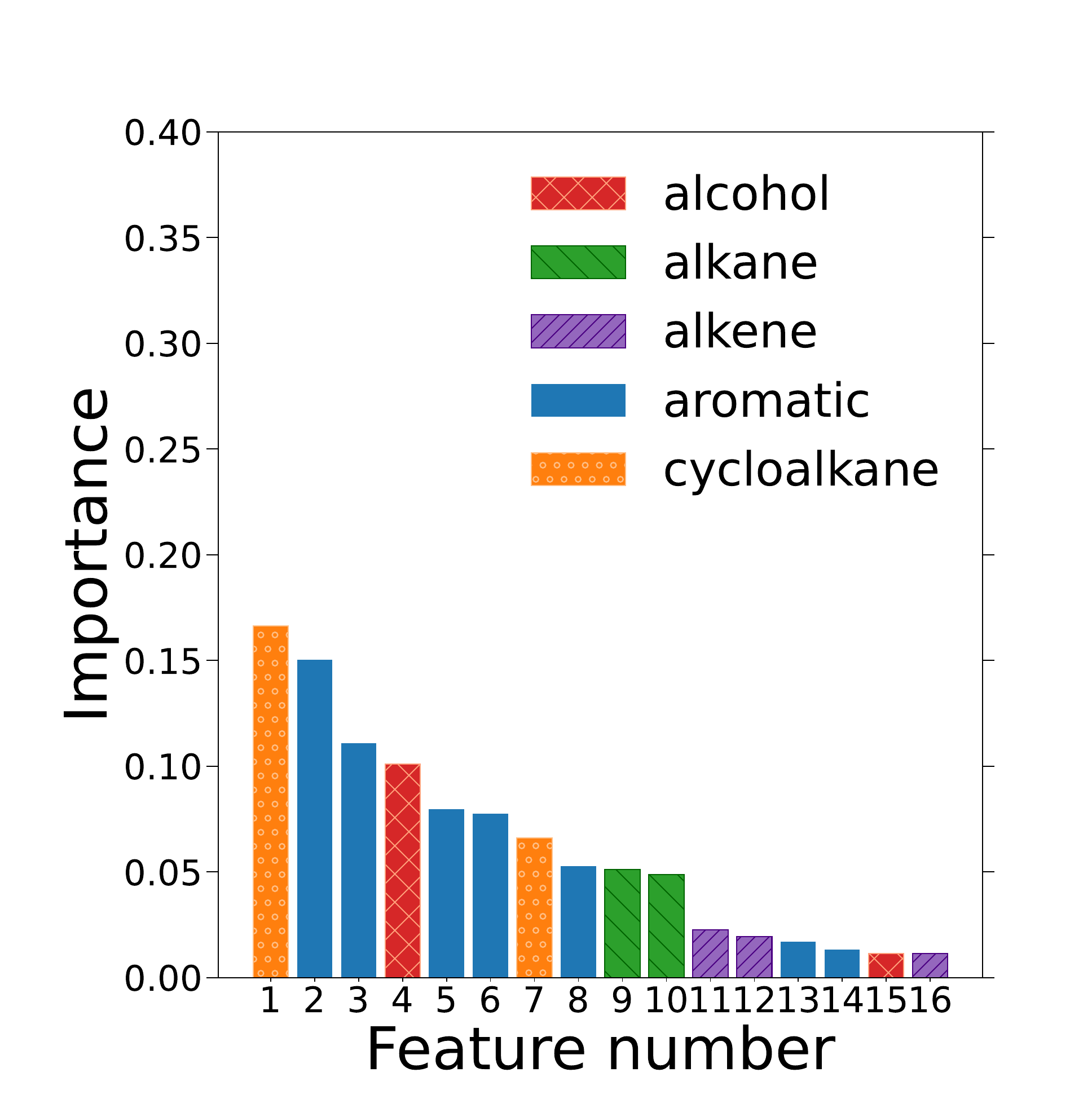}
        \caption{final boiling point}
        \label{fig:bp_imp}
    \end{subfigure}
    \begin{subfigure}[b]{0.33\textwidth}
        \includegraphics[width=\textwidth]{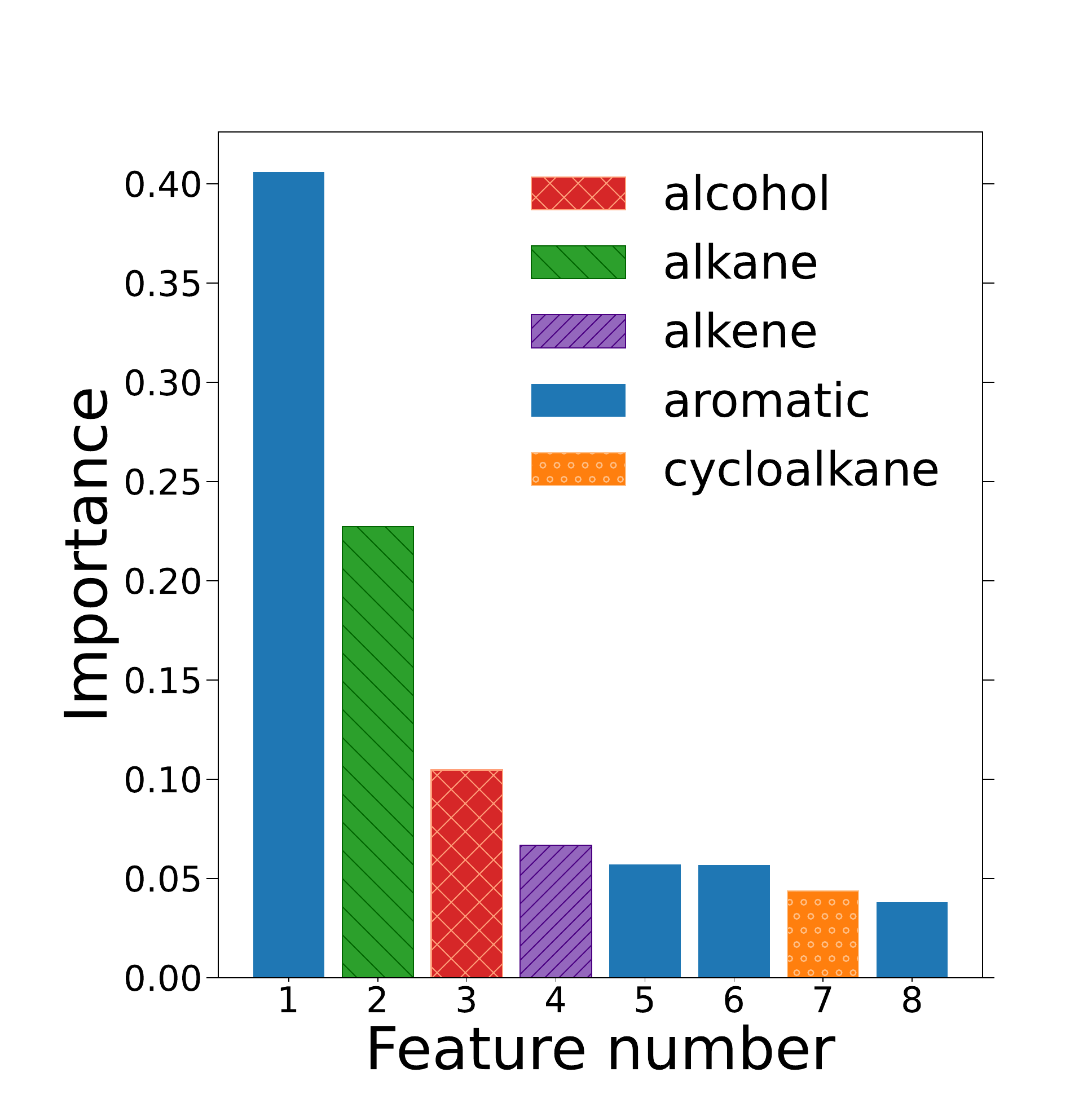}
        \caption{flash point}
        \label{fig:fp_imp}
    \end{subfigure}
    \begin{subfigure}{0.33\textwidth}
    \centering
        \includegraphics[width=\textwidth]{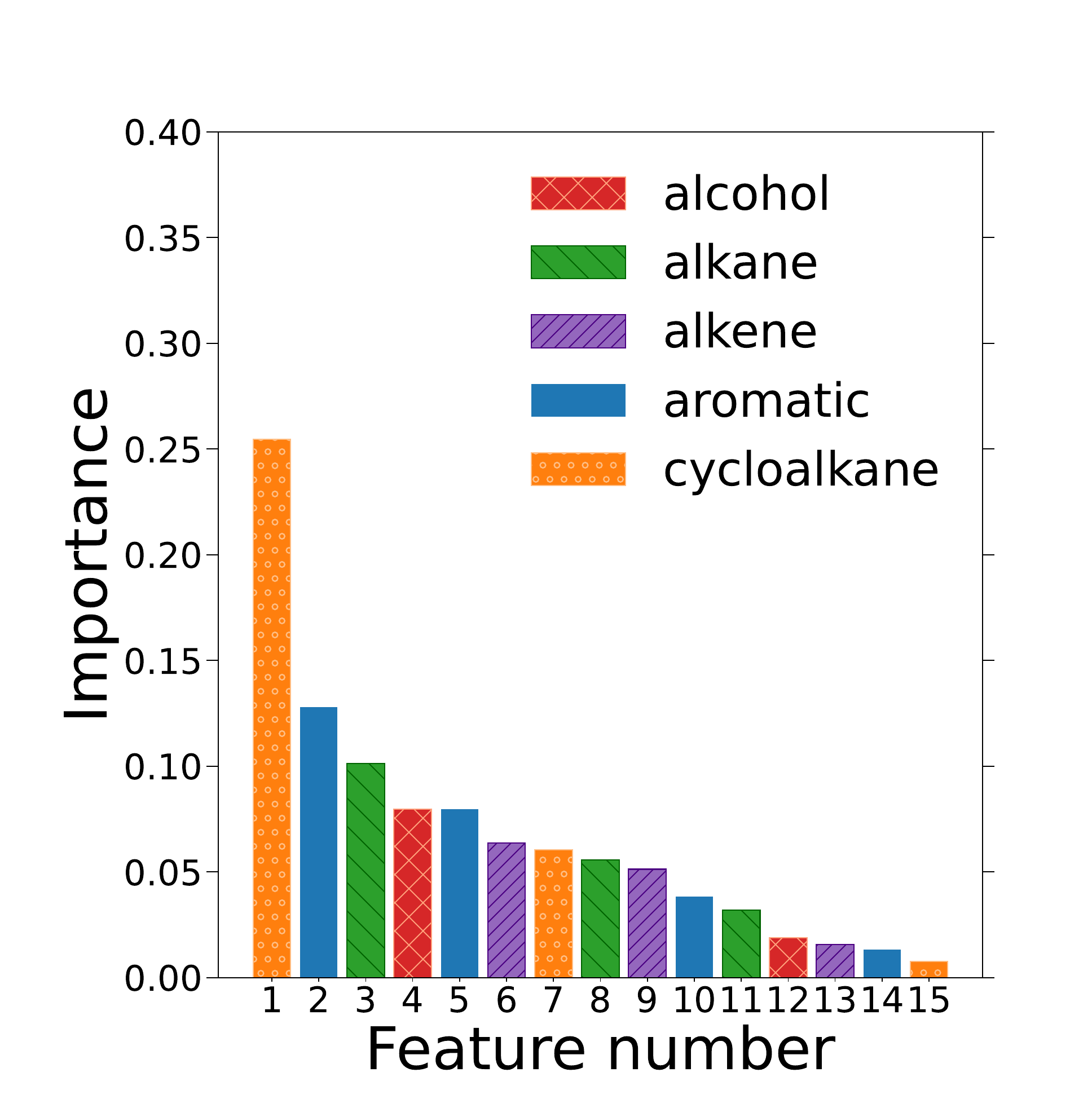}
        \caption{freezing point}
        \label{fig:mp_imp}
    \end{subfigure}

    \hspace{0.1515\textwidth}
        \begin{subfigure}{0.33\textwidth}
        \centering
        \includegraphics[width=\textwidth]{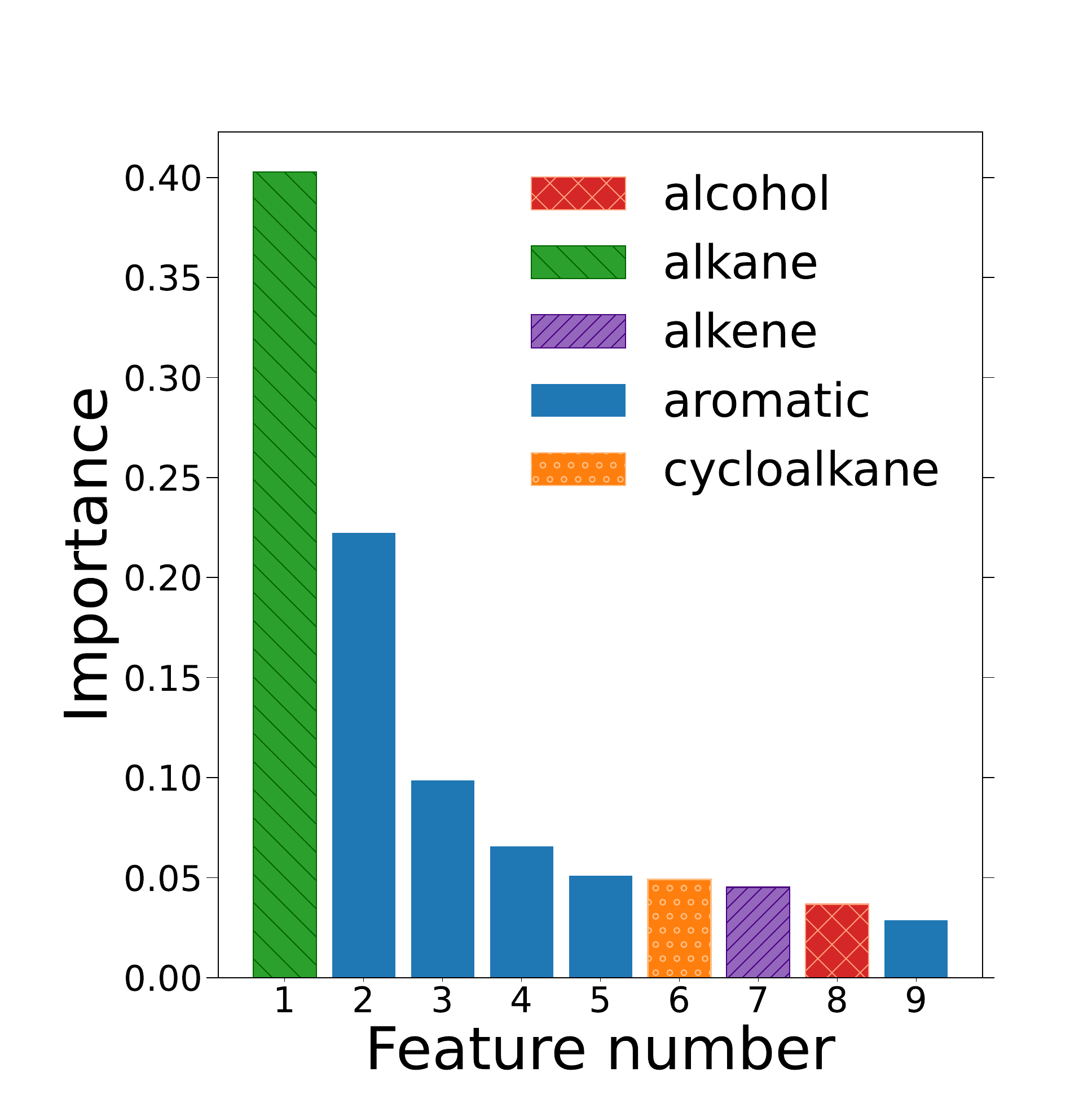}
        \caption{density at \qty{15}{\degreeCelsius}}
        \label{fig:den_imp}
    \end{subfigure}
    \begin{subfigure}{0.33\textwidth}
    \centering
        \includegraphics[width=\textwidth]{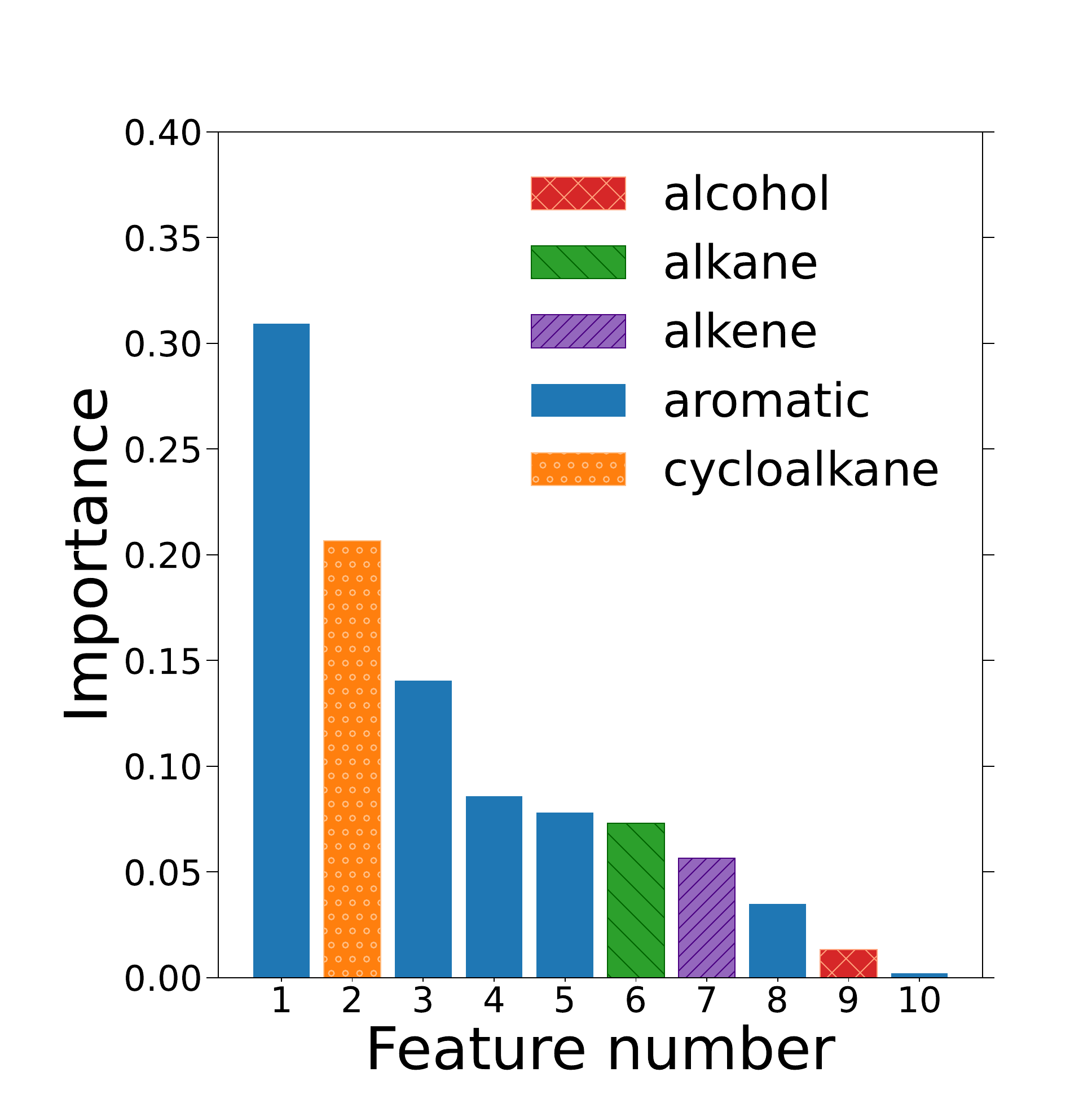}
        \caption{kinematic viscosity at \qty{-20}{\degreeCelsius}}
        \label{fig:vic_imp}
    \end{subfigure}

    \caption{Model feature importances for all models.}
    \label{fig:all_importances}
\end{figure}

In addition to creating accurate predictive models, this method also provides interpretable features that highlight possible attributes most affecting each property. 
For example, Figure~\ref{fig:all_importances} shows the types of features used in each property model and their importances.
Features are grouped by compound class using identified spectra bands and regions associated with alcohols, alkanes, alkenes, aromatics, and cylcoalkanes.
Feature importance is calculated by the model and assesses the reduction of impurity by each feature (i.e., how well the data is split between trees)~\cite{rf_breiman}.
The features are ranked in order of importance, with Feature 1 ranked highest.

Figure~\ref{fig:all_importances} shows that most models have between one and three dominant features, with the exception of final boiling point, which contains a more diverse set of equally important features.
The top two features in the flash point model (Figure~\ref{fig:all_importances}b) contain spectral peaks similar to aromatics and alkanes with an aggregated feature importance of about 0.64.
For the freezing point model (Figure~\ref{fig:all_importances}c), the highest-ranked feature contains cycloalkane attributes and has an importance of 0.25, about twice the importance of the second ranked feature.
The density model's highest ranking features contain alkane and aromatic characteristics, with the alkane feature having an importance of 0.4 (see Figure~\ref{fig:all_importances}d).
The top three features in the viscosity model (Figure~\ref{fig:all_importances}e) have an aggregated feature importance of over 0.65 and contain spectra peaks similar to aromatics and cycloalkanes.

\begin{figure}[!b]
\centering
\begin{minipage}{.5\textwidth}
  \centering
  \includegraphics[width=\textwidth,height=\textwidth,keepaspectratio]{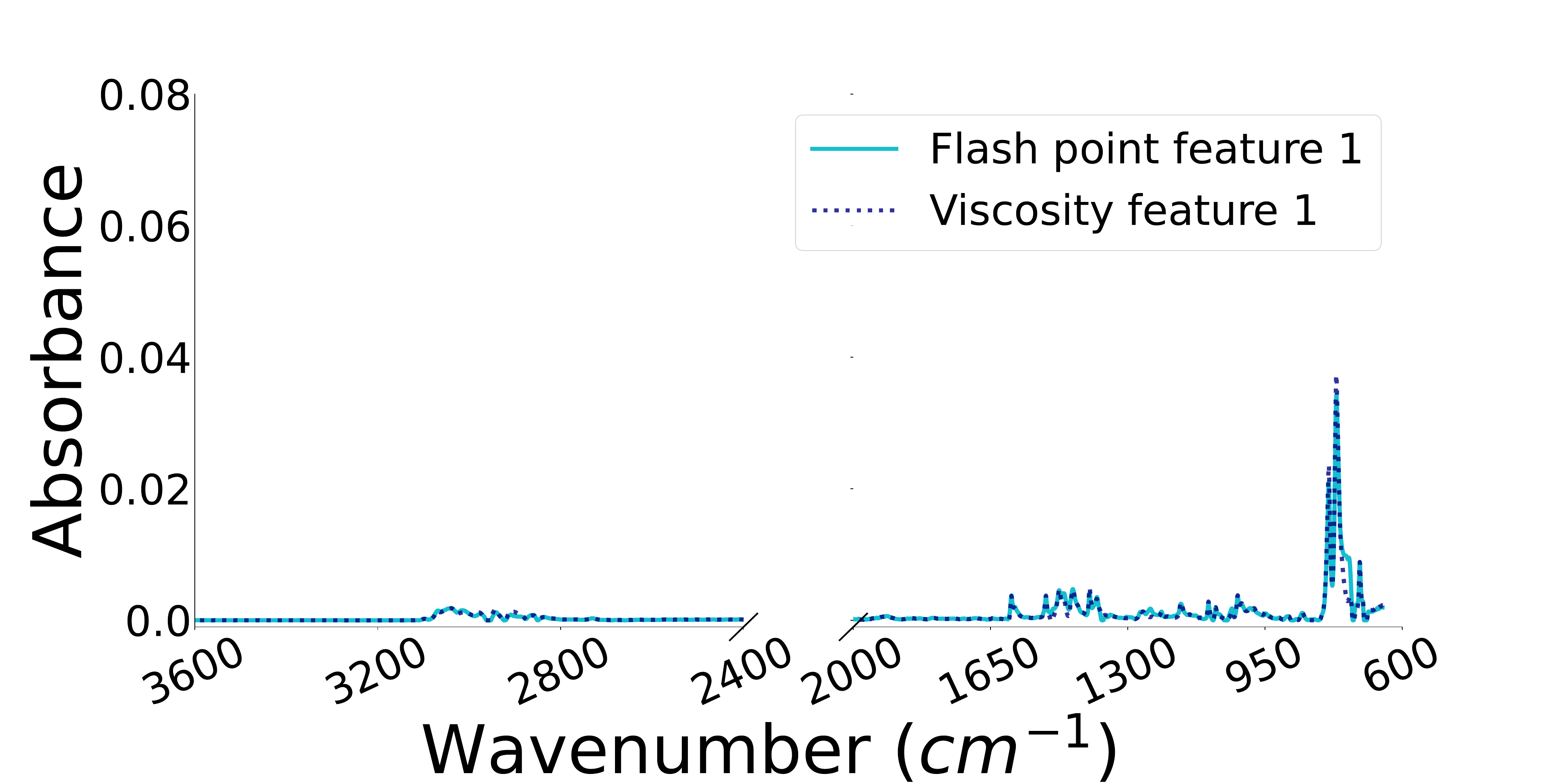}
  \captionsetup{justification=centering}
  \captionof{figure}{Highest ranked aromatic features for\\ flash point and viscosity models.}
  \label{fig:arom_fp_vic}
\end{minipage}%
\begin{minipage}{.5\textwidth}
  \centering
  \includegraphics[width=\textwidth,height=\textwidth,keepaspectratio]{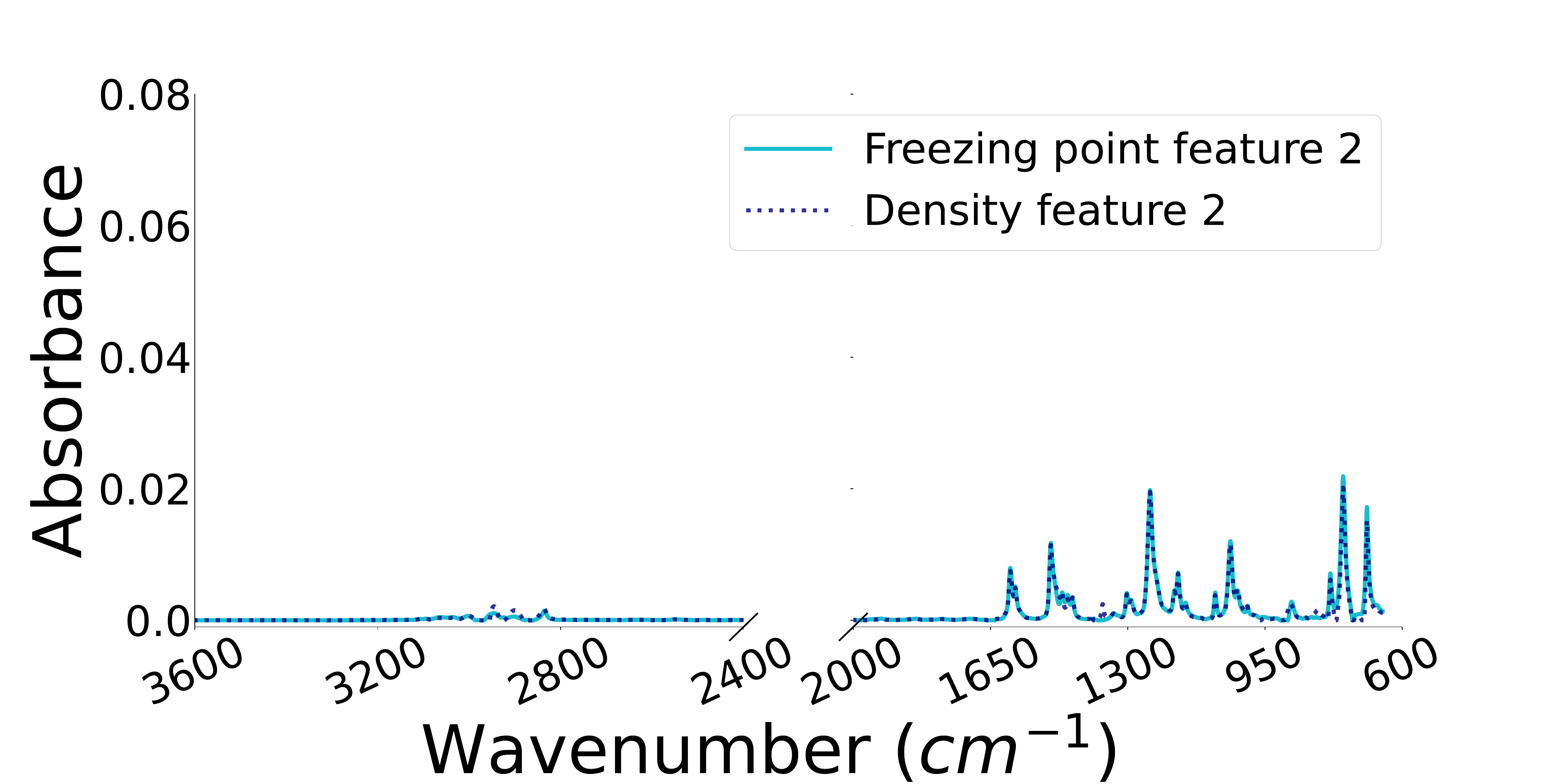}
  \captionsetup{justification=centering}
  \captionof{figure}{Highest ranked aromatic features for\\ freezing point and density models.}
  \label{fig:arom_mp_den}
\end{minipage}
\end{figure}

For all of the models, features with aromatic characteristics are ranked either first or second.
Interestingly, Figure~\ref{fig:arom_fp_vic} shows that the top aromatic feature in the flash point and viscosity models are almost indistinguishable, exhibiting a prominent peak near \qty{770}{\cm^{-1}}, characteristic to terminal ethyl groups~\cite{Scheuermann2017}. 
Several other out-of-plane \ce{C-H} bending bands associated with aromatics can also be seen in the region \qtyrange{900}{675}{\cm^{-1}}~\cite{ftir_chem350}.
Smaller peaks representing \ce{C-C} in-ring stretching are located over \qtyrange{1500}{1400}{\cm^{-1}} and \qtyrange{1600}{1585}{\cm^{-1}}~\cite{ftir_chem350}.

The top aromatic features match between the freezing point and density models, which are also the second-highest ranking features in both models (see Figure~\ref{fig:arom_mp_den}).
Both features contain prominent peaks between \qtyrange{800}{650}{\cm^{-1}} representing out-of-plane \ce{C-H} bands~\cite{ftir_chem350}.
Other peaks near \qty{1600}{\cm^{-1}} and from \qtyrange{1500}{1400}{\cm^{-1}} represent carbon-carbon stretching vibrations in the aromatic rings~\cite{ftir_chem350}.
These features also include a prominent peak around \qty{1244}{\cm^{-1}}, likely representing \ce{C-C} stretching from terminal tert-butyl groups~\cite{Scheuermann2017}.

\begin{figure}[!t]
\centering
\begin{minipage}{.5\textwidth}
  \centering
  \includegraphics[width=\textwidth,height=\textwidth,keepaspectratio]{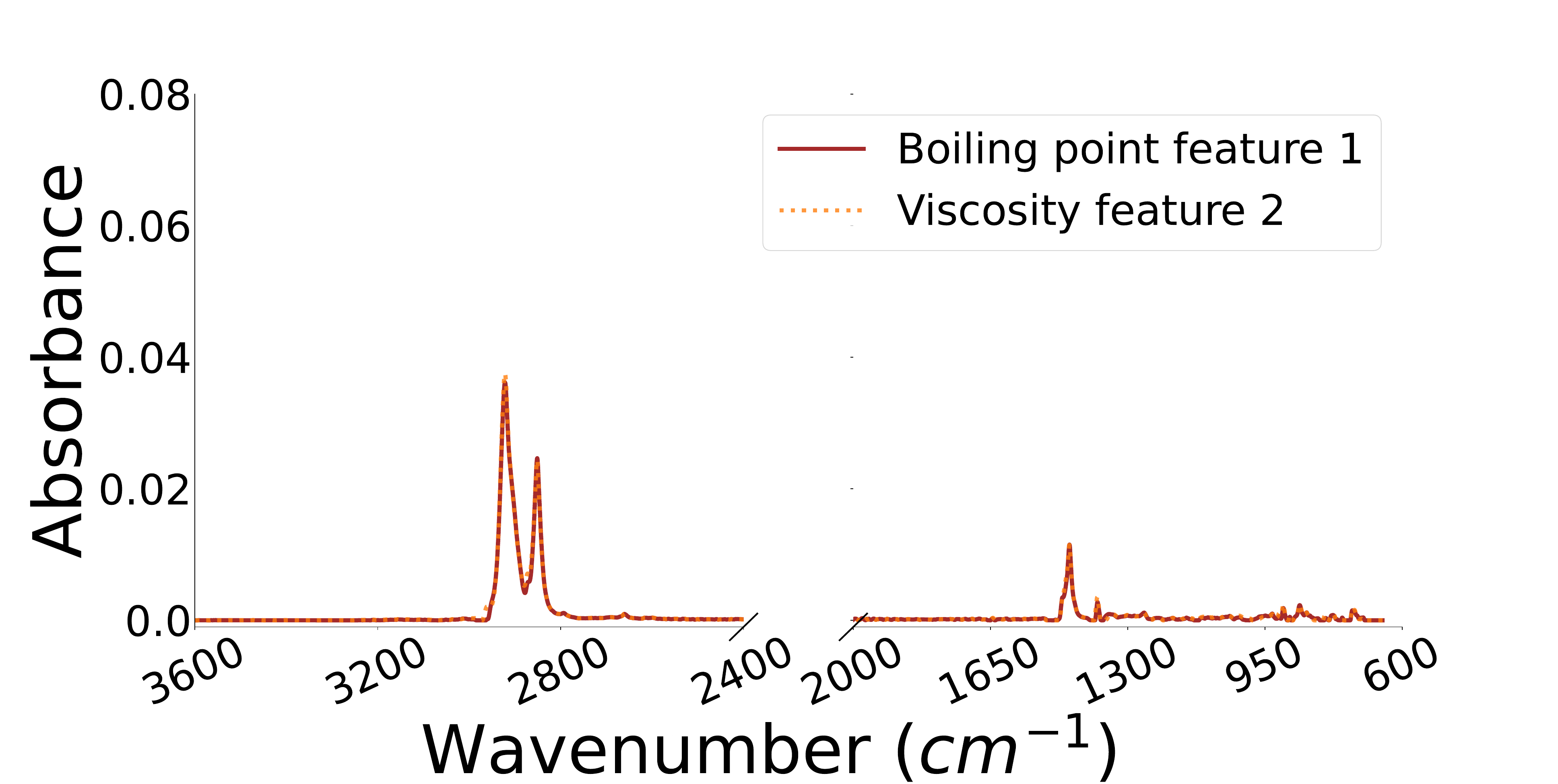}
  \captionsetup{justification=centering}
  \captionof{figure}{Highest ranked cycloalkane features for\\ final boiling point and viscosity models.}
  \label{fig:cyclo_bp_vic}
\end{minipage}%
\begin{minipage}{.5\textwidth}
  \centering
  \includegraphics[width=\textwidth,height=\textwidth,keepaspectratio]{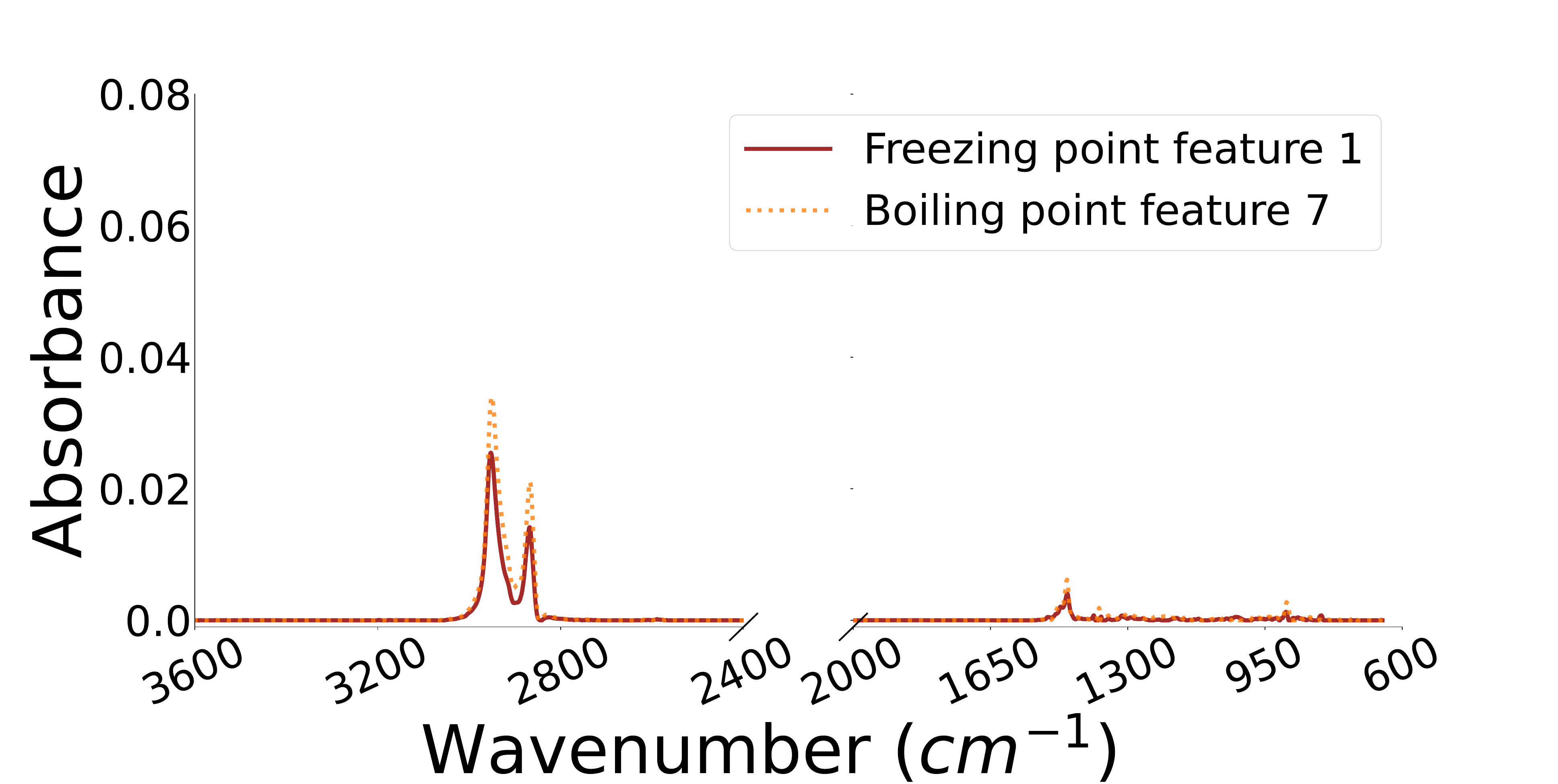}
  \captionsetup{justification=centering}
  \captionof{figure}{Highest ranked cycloalkane feature in the\\ freezing point model and second-highest ranked cycloalkane feature in the final boiling point model.}
  \label{fig:cyclo_bp_mp}
\end{minipage}
\end{figure}

Features containing cycloalkane attributes are ranked either first or second in the final boiling point, freezing point, and viscosity models. 
Similar to the aromatic features, the top cycloalkane features in the final boiling point and viscosity models exhibit nearly identical peaks (see Figure~\ref{fig:cyclo_bp_vic}). 
The features exhibit two peaks near \qty{2850}{\cm^{-1}} and \qty{2920}{\cm^{-1}} representing asymmetric and symmetric stretching vibrations of \ce{CH2} groups, as well as a smaller peak near \qty{1450}{\cm^{-1}} representing \ce{C-H} bending or scissoring~\cite{puhan_2023,ftir_chem350}. 
The peaks captured by these features closely match spectral data from cyclohexane-based compounds.
In addition, the top cycloalkane feature in the freezing point model closely resembles the second cycloalkane feature in the final boiling point model (see Figure~\ref{fig:cyclo_bp_mp}).
Both of these features contain a peak between \qty{1470}{\cm^{-1}} and \qty{1450}{\cm^{-1}} attributed to \ce{C-H} bending or scissoring, and \ce{C-H} stretching peaks near \qty{3000}{\cm^{-1}}~\cite{ftir_chem350}.

\begin{figure}[!t]
\centering
\includegraphics[width=0.5\textwidth,height=0.5\textwidth,keepaspectratio]{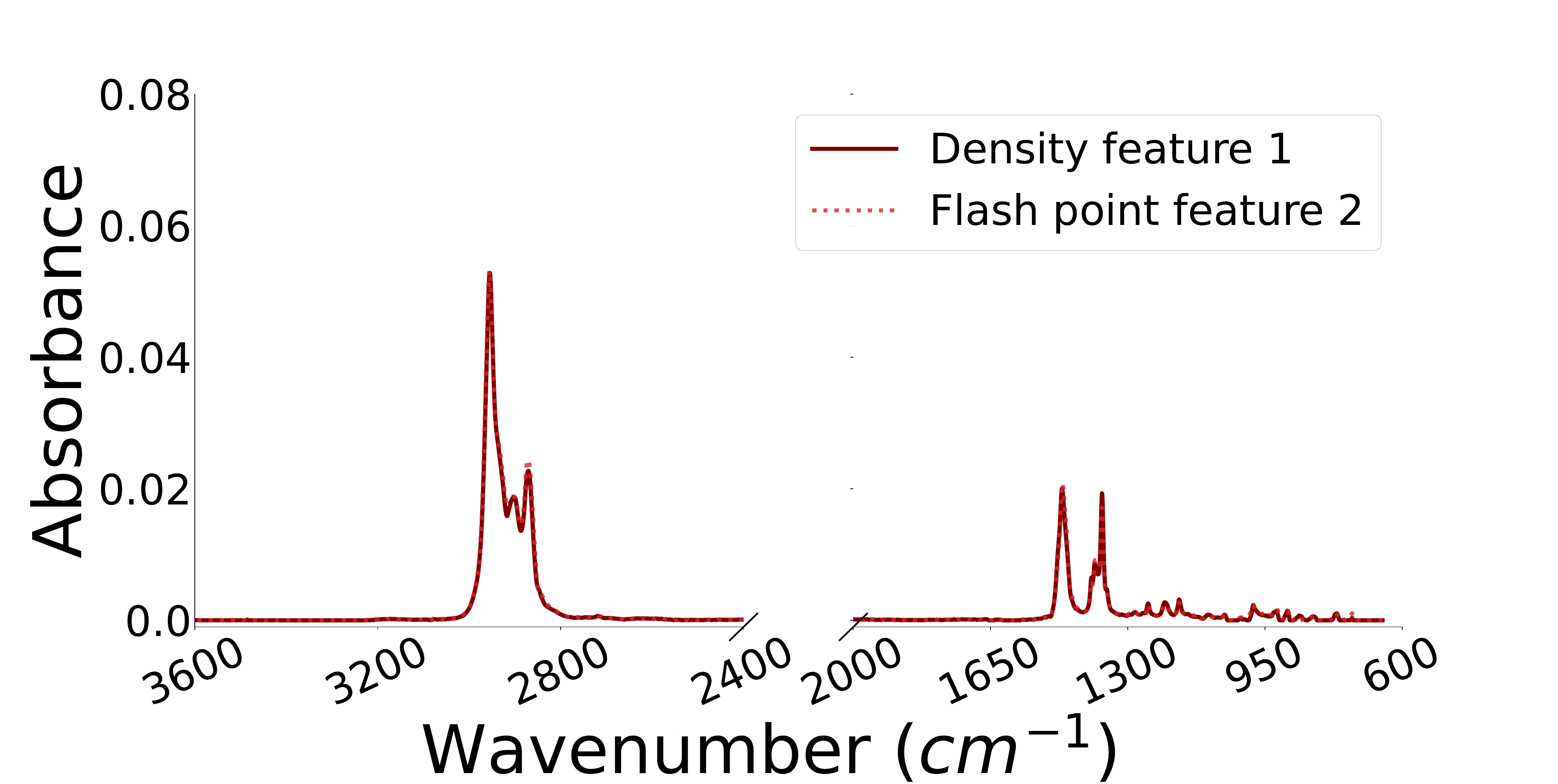}
\caption{Highest ranked alkane features for the density and flash point models.} 
\label{fig:alka_den_fp}
\end{figure}

Alkane-like features also play an important role in the density and flash point models, ranking first and second, respectively.
The high importance of alkanes for predicting density could be due to alkanes generally having lower densities.
The model indicates that alkane-like features in a spectra would result in lower predicted density.
Figure~\ref{fig:alka_den_fp} shows the top features with alkane characteristics for both models are almost indistinguishable.
The features exhibit \ce{C-H} stretching between \qty{3000}{\cm^{-1}} and \qty{2850}{\cm^{-1}}, and \ce{C-H} bending or scissoring between \qtyrange{1470}{1450}{\cm^{-1}}~\cite{ftir_chem350}.
In addition, the features contain \ce{C-H} rocking bands attributed to methyl groups between \qtyrange{1370}{1360}{\cm^{-1}}~\cite{ftir_chem350}.
The features lack \ce{C-H} rocking bands seen in long-chain n-alkanes from \qtyrange{725}{720}{\cm^{-1}} and most likely represent shorter n-alkanes or branched alkanes such as isooctane~\cite{ftir_chem350}.
Further details for other model features used in each property can be found in Section S-4 of the SI.

\section{Conclusions}\label{concl}

This research provides a structured method for creating accurate and interpretable property prediction models using liquid-phase \gls{FTIR} spectra of neat molecules, aviation fuels, and blends.
Liquid-phase \gls{FTIR} spectra can can be collected rapidly and consistently, characterizing compounds with only a few drops of a sample.
The method uses \gls{NMF} to deconstruct \gls{FTIR} spectra into fundamental building blocks that are used as features to train machine learning models.
Deconstruction occurs intuitively, preserving inherent characteristics of spectral data by leveraging the nonnegativity constraint.
Furthermore, this method facilitates interpretation for discovering new relationships between compositional elements of a fuel, such as functional groups or chemical classes, and its properties.

To demonstrate the method, five ensemble models were developed to predict final boiling point, flash point, freezing point, density at \qty{15}{\degreeCelsius}, and kinematic viscosity at \qty{-20}{\degreeCelsius}.
The samples used to train and test the models included neat molecules, aviation fuels, and blends.
The number of neat molecules, aviation fuels, and blends in the property databases ranged from 31 (kinematic viscosity) to 72 (final boiling point).
The models' performance aligns with published literature, having test mean absolute errors of \qty{17.7}{\kelvin}, \qty{9.6}{\kelvin}, \qty{8.9}{\kelvin}, \qty{22.2}{\kg\per\m^3}, and \qty{0.45}{\mm^2\per\s} for final boiling point, flash point, freezing point, density, and kinematic viscosity, respectively.
Because some of the experimental datasets were discretized due to the large ranges in property values, more experimental property data would likely increase the accuracy of the models.

The results also showed that this method provides interpretable features that enable the investigation of chemical classes, functional groups, and bonds that contribute most to property prediction.
Specifically, all models have features with aromatic characteristics ranked either first or second, with flash point and viscosity models sharing top aromatic features that are almost indistinguishable.
The top aromatic features for the flash point and viscosity models exhibit a prominent peak characteristic to terminal ethyl groups, several other out-of-plane \ce{C-H} bending bands associated with aromatics, and smaller peaks representing \ce{C-C} in-ring stretching.
Top aromatic features for freezing point and density models also show nearly identical characteristics, including prominent peaks representing out-of-plane \ce{C-H} bands, carbon-carbon stretching vibrations in the aromatic rings, and \ce{C-C} stretching from terminal tert-butyl groups.
Additional important features for final boiling point, freezing point, and viscosity models include cycloalkane characteristics, while density and flash point models highly rank features with alkane characteristics.

Overall, the method is able to consistently predict properties of neat molecules, aviation fuels, and blends, while providing interpretable features that enable scientific discovery and help accelerate fuel research. 
To support research and development, the models and data have been integrated into an existing webtool, located at \url{feedstock-to-function.lbl.gov}.
To improve and expand the method, future work should explore additional analytical techniques with a substantial amount of spectral and experimental property data.

    \section{CRediT authorship contribution statement}
    
    \textbf{Ana E. Comesana}: Conceptualization, Methodology, Software, Validation, Data curation, Writing – original draft, Writing – review \& editing, Visualization. \textbf{Sharon S. Chen}: Methodology, Data curation, Writing – review \& editing. \textbf{Kyle E. Niemeyer}: Conceptualization, Writing – review \& editing, Supervision, Funding acquisition. \textbf{Vi H. Rapp}: Conceptualization, Methodology, Validation, Writing – original draft, Writing – review \& editing, Supervision, Funding acquisition.

    \section{Acknowledgements}\label{ackn}
    This work was supported by the Bioenergy Technologies Office of the U.S. Department of Energy through Contract DE-AC02-05CH11231 between Lawrence Berkeley National Laboratory and the US Department of Energy. 
    
    Work at the Molecular Foundry was supported by the Office of Science, Office of Basic Energy Sciences, of the U.S. Department of Energy under Contract No. DE-AC02-05CH11231.
    
    This research also used resources of the Oak Ridge Leadership Computing Facility, which is a Department of Energy Office of Science User Facility supported under Contract DE-AC05-00OR22725. The United States Government retains and the publisher, by accepting the article for publication, acknowledges that the United States Government retains a nonexclusive, paidup, irrevocable, world-wide license to publish or reproduce the published form of this manuscript, or allow others to do so, for United States Government purposes.
    
    The authors would also like to acknowledge Ben Harvey, Paul Wrzesinski, and Milissa Flake for supplying aviation fuels and compounds critical to the success of this project; Josh Heyne for providing a wealth of data, insights, and feedback; Jaspreet Singh and Mark McMahon for their support with interpreting FTIR spectra data, and for helping us obtain, operate, and calibrate equipment for FTIR spectra measurements; Liana Klivansky for her support with operating and interpreting data from the FTIR spectrometer; Cody Bastien for ensuring the fuel property certification measurements were repeatable and robust; and Michael Golden for his support with gaining access to published literature and data.
    
    \nolinenumbers
    \biboptions{sort&compress}
    \bibliography{mybibfile}

\end{document}